\documentstyle[preprint,prl,aps]{revtex}
\begin{document}
\draft
\title{Decoherence in Ion Trap Quantum Computers}
\author{Anupam Garg\cite{emad}}
\address{Department of Physics and Astronomy,\\
         Northwestern University, Evanston, Illinois 60208}
\date{\today}
\maketitle

\begin{abstract}
The {\it intrinsic} decoherence from vibrational
coupling of the ions in the Cirac-Zoller quantum computer [Phys. Rev.
Lett. {\bf 74}, 4091 (1995)] is considered. Starting from a
state in which the vibrational modes are at a temperature $T$,
and each ion is
in a superposition of an excited and a ground state, an adiabatic
approximation
is used to find the inclusive probability $P(t)$ for the ions to evolve as
they would without the vibrations, and for the vibrational modes to evolve
into any final state. An analytic form is found for $P(t)$ at $T=0$, and the
decoherence time is found for all $T$. The decoherence is found to be quite
small, even for 1000 ions.
\end {abstract}


\narrowtext

Quantum computers (QC) are (as yet hypothetical) devices
with states that are
quantal in nature, and which perform calculations by unitary
transformations on these states \cite{Ben,AE,JSN}. The
linearity of the superposition principle leads to an inbuilt massive
parallelism: a computer with $N$ two-state elements can operate on $2^N$
states simultaneously. This parallelism underlies Shor's recent
algorithm \cite{Sh} for factorizing a composite number of order $2^L$
in $\sim L^3$ steps on a QC. The best known classical algorithm 
takes $\sim\exp[c(L) L^{1/3}]$ steps with $c(L)\sim (\ln L)^{2/3}$.
The potential for other quantum algorithms is clearly exciting,
as is the emergence of
a new paradigm for computation itself.

It is obvious that
maintaining perfect phase coherence among all the states of a QC
is a daunting task, not to mention getting these states to
evolve in the desired fashion in the first place \cite{Lan}. 
An imaginative proposal for a QC by Cirac and Zoller \cite{CZ}
(CZ) seems promising in addressing these
problems \cite{Win,Kim}. It utilizes a string of $N$ identical
ions in a linear
Paul trap \cite{Rai}, with each ion separately addressable by a laser.
Two internal states of each ion, $|e\rangle$ and $|g\rangle$,
are used for the QC, along with the center-of-mass (CM) axial
vibrational mode of the entire array. A program is implemented as a
specified sequence of ($\pi/2$, $\pi$, etc.) pulses that drive 
$|e\rangle \leftrightarrow |g\rangle$
transitions on any given ion, along with pulses detuned by the CM
frequency that enable coupled transitions between any pair of ions.

Two types of decoherence should be distinguished in the CZ (indeed,
any) QC. The first is technical, due, e.g., to imperfect phase
locking, mistuning of lasers,
errors in timing and duration of pulses, and overlooked
perturbations in the Hamiltonian. The second kind is intrinsic, and
arises from coupling of the computationally useful to the undesirable
bath degrees of freedom.  Although the technical problems
alone render the pursuit of a QC a fool's quest in many
people's eyes, intrinsic decoherence sets basic
limits on the capabilties of a QC. It is with this motivation that we
study intrinsic decoherence in the CZ QC \cite{Un}.

We take as our bath the vibrations
of the ions, which we treat as undamped harmonic oscillators.
Damping can be included if necessary \cite{lat}.
Radiative decoherence is accounted for very simply by demanding
that any computation take less time than $\tau_{\rm sp}/N$, where
$\tau_{\rm sp}$ is the spontaneous $|e\rangle \to |g\rangle$ decay time for
one ion. It clearly pays to have as large a $\tau_{\rm sp}$ as possible,
by working with E1 forbidden transitions \cite{CZ}, or with
hyperfine sublevels of the ground ionic multiplet \cite{Win}.
The total Hamiltonian minus the driving lasers
can be generally written as (setting $\hbar =1$)
\begin{equation}
H = {1\over 2} \sum_i \omega_0 \sigma_{iz} +
    \sum_{\mu} ({p^2_{\mu} \over {2m}} +
          {1 \over 2} m \omega^2_{\mu} q^2_{\mu} ) +
    \sum_{i,\mu} \sigma_{i\perp}\cdot {{\bf c}}_{i \mu} q_{\mu}.
\label{Htot}
\end{equation}
Here, the $\sigma$'s are equivalent Pauli spin operators in the
$\{ |e\rangle, |g\rangle \} $ space, 
$\sigma_{\perp} = (\sigma_x, \sigma_y, 0)$,
$q_{\mu}$ and $p_{\mu}$ are the
vibrational normal mode coordinates and momenta, and $m$ is the
mass of each ion. We shall refer to the three terms in
Eq.~(\ref{Htot}) as $H_i$, $H_{\rm nm}$, and $H'$ respectively.
The ${{\bf c}}_{i \mu}$ are calculable functions (see below)
of the ionic
transition matrix elements and equilibrium ion positions, which
we assume are such that there
is no $|e\rangle \leftrightarrow |g\rangle$ transition term in $H$
in equilibrium. It is key to successful operation of the CZ QC
that the vibrations be cooled to nearly zero temperature, and that the
frequencies $\omega_{\mu}$ and the couplings ${\bf c}_{i\mu}$
be small. The approximations of this paper require that
$\omega_0 \gg \omega_{\mu},
\langle \sum_{\mu} c_{i\mu}q_{\mu}\rangle$, and
$\sum_{\mu}\langle c_{i\mu} q_{\mu} \rangle^2/\omega_0
               \ll \omega_{\mu}$,
which as we shall see, can be satisfied comfortably.
 
Let us now study the effects of the bath on the simplest
computation of all, i.e., just waiting. We include the CM 
vibrational mode in
the bath for simplicity in this note, as this is not
expected to change the result qualitatively. Suppose that initially,
the bath is described by a density matrix $\rho(Q, Q')$ ($Q$ denotes all
the $q_{\mu}$ collectively), and the ions are in some state
$|{\rm in}\rangle$. The system is not driven by any lasers, and simply
sits for a time $t$. What is the probability $P(t)$ of finding the ions in
the final internal state
$|{\rm fin}\rangle \equiv \exp(-iH_i t) |{\rm in} \rangle$
that one would get in the absence of the bath,
and the bath in any state whatsoever? For the state
$|{\rm in}\rangle$, we take
\begin{equation}
|{\rm in}\rangle = \prod_i 2^{-1/2}(|+\rangle_i + |-\rangle_i),
\label{instate}
\end{equation}
where $\sigma_{iz} |\pm\rangle_i = \pm |\pm\rangle_i$. This state
is illustrative of the complex superpositions of computational basis
states that give QC's their parallelism.
Since Eq.~(\ref{Htot})
describes a finite, closed system, $P(t) \not\to 0$ as $t \to \infty$,
but we expect that $P(t)$ will drop close to zero at some time $\tau_d$,
after which it will fluctuate with small amplitude \cite{recur}.
The time $\tau_d$ limits the longest computation
that can be done with the CZ QC
(if $\tau_d < \tau_{\rm sp}/N$). The coherence time is expected
to decrease when transitions are driven by the lasers, and can also
be estimated \cite{lat}. 

To evaluate $P(t)$, we write the reduced bath density
matrix propagator as a double path integral
\widetext
\begin{equation}
J(Q_f,Q_f';Q,Q') = \int_{Q'}^{Q'_f}\!\int_Q^{Q_f} [dQ][dQ']
                   e^{i(S_0[Q(t)] - S_0[Q'(t)])} A[Q(t)] A^*[Q'(t)],
\label{prop}
\end{equation}
\narrowtext
where $S_0$ is the action for the bath alone, and
\begin{equation}
A[Q(t)] = \left \langle {\rm fin} \left| {\rm T}
          \exp{\left(- i
          \int_0^t[H_i + H'(Q(t'))]\, dt' \right)} \right|
               {\rm in} \right \rangle.
\label{ampl}
\end{equation}
In terms of $J$, $P(t)$ is given by
\begin{equation}
P(t) = \int\!\!\int\!\!\int dQ_f\,dQ\,dQ'\,
                J(Q_f,Q_f;Q,Q') \,\rho(Q,Q').
\label{Qint}
\end{equation}
It now does no good to integrate out the oscillators.
Instead, we exploit the fact that
$\omega_0 \gg \omega_{\mu}$---typically
$\omega_0 \sim 10^{15}\, {\rm s}^{-1}$, and
$\omega_{\mu} \sim 10^7\, {\rm s}^{-1}$---to integrate out
{\it the spins}. It is easily
seen that $A[Q(t)]$ factorizes into
$\prod_i A_i[Q(t)]$. The {\it i\/}th spin experiences a field
\mbox{\boldmath $\omega$}$_{i\perp} = 
2\sum_{\mu}{{\bf c}}_{i\mu} q_{\mu}(t)$
in the {\it x-y} plane, with magnitude
$\omega_{i\perp} \ll \omega_0$ and
$|d\ln\omega_{i\perp}/dt| \ll \omega_0$, as we may safely assume that
Eq.~(\ref{prop}) is dominated by paths $Q(t)$ varying slowly on 
the $\omega_0^{-1}$ time scale. This permits us to evaluate
$A_i[Q(t)]$ using an adiabatic approximation.
We may further take the instantaneous
precession axis of the spin as $\hat{\bf z}$ for all $t$ with
negligible error (of order $\omega_{i\perp}/\omega_0$). It is
far more important to approximate the phase well. The instantaneous
energies of the states $|\pm\rangle_i$ are given by
$\pm (\omega_0 +\omega_{i\perp}^2/2\omega_0)$ to relative order 
$\omega_{i\perp}/\omega_0$. We thus obtain
$A_i = \cos\Phi_i(t)$ with
\begin{equation}
\Phi_i(t) = \int_0^t dt'\, \omega_{i\perp}^2(t')/4\omega_0.
\label{phase}
\end{equation}
Equation~(\ref{phase}) holds for
$t < O(8\omega_0^3/\langle{\dot\omega}^2_{i\perp} \rangle)$, where
$\langle\rangle$ denotes an average value, and the dot denotes $d/dt$.
(The results obtained below imply that this time
scale exceeds $\tau_d$.)
We have also omitted a negligible Berry phase term \cite{lat}.

It now pays to rearrange the expression for $P(t)$.
First, we write $\cos\Phi_i$ as a sum of $e^{\pm i\Phi_i}$,
and substitute the
resulting expression for $A[Q(t)]$ in Eq.~(\ref{prop}). This yields,
\begin{equation}
J(Q_f,Q'_f; Q, Q') = {1\over 2^{2N}} \sum_{\{s\}, \{s'\}}
                      K_{\{s\}}(Q_f,Q) K_{\{s'\}}^*(Q'_f,Q'),
\label{propK}
\end{equation}
with
\widetext
\begin{equation}
K_{\{s\}}(Q_f,Q) = \int_Q^{Q_f} [dQ]\,
          \exp\left( iS_0[Q(t)] + i \sum_{i,\mu,\nu} s_i u^i_{\mu\nu}
          \int_0^t q_{\mu}(t) q_{\nu}(t)\, dt \right).
\label{Ker}
\end{equation}
\narrowtext
In Eqs.~(\ref{propK}) and (\ref{Ker}),
$\{s\} = (s_1,s_2,\ldots,s_N)$, $\{s'\}$ is similarly given,
each $s_i = \pm 1$ is an Ising-like variable, and
$u^i_{\mu\nu} =  {\bf c}_{i\mu}\!\cdot\! {\bf c}_{i\nu}/\omega_0$.
Next we define the following combination of propagators,
\begin{equation}
R_{\{s,s'\}} (Q, Q') = \int dQ_f\,  K_{\{s\}} (Q_f, Q)
                K_{\{s'\}}^* (Q_f,Q'),
\label{difprop}
\end{equation}
in terms of which $P(t)$ can be written as [see
Eqs.~(\ref{Qint}) and (\ref{propK})]
\begin{equation}
P(t) = {1\over 2^{2N}}\sum_{\{s\},\{s'\}}
       \int\!\!\int dQ\,dQ'\, R_{\{s,s'\}} (Q,Q') \rho(Q,Q').
\label{Qint2}
\end{equation}

It is apparent that $K_{\{s\}}$ is the propagator for a set
of coupled harmonic oscillators, described by a Hamiltonian
that depends on the Ising congiguration $\{s\}$:
\begin{eqnarray}
H_{\rm nm}(\{s\}) &=& \sum_\mu {p_{\mu}^2 \over 2m} +
         {1\over 2}\sum_{\mu,\nu} q_{\mu}(\Omega^2)_{\mu\nu} q_{\nu},
\label{Hs}\\
(\Omega^2)_{\mu\nu} &=&  \omega^2_{\mu}\delta_{\mu\nu} 
                      - 2\sum_{i}s_i\,u^i_{\mu\nu}.
\label{Om}
\end{eqnarray}
We can thus write $R_{\{s,s'\}}$ alternatively as
\begin{equation}
R_{\{s,s'\}}(Q, Q') = \langle Q' |
    e^{iH_{\rm nm}(\{s'\})t}
    e^{-iH_{\rm nm}(\{s\})t} | Q\rangle. 
\label{Rham}
\end{equation}
Since Eq.~(\ref{Rham}) only involves harmonic oscillators,
we can evaluate it exactly by reverting to path integrals.
The exact answer involves trigonometric functions
and determinants of the matrices $\Omega(\{s\})$ and $\Omega(\{s'\})$
and is of limited use because of the
remaining sum on the $s$'s. To make further progress, we employ
an approximation in the same spirit as that used to obtain the spin
transition amplitude $A_i(Q[t])$. Namely, accuracy in the normal
mode frequencies is much more important than in the normal modes
themselves. Errors in the former lead to errors in $K_{\{s\}}$
and $R_{\{s,s'\}}$
that grow with time, while errors in the latter do not. We therefore
treat the second term in Eq.~(\ref{Om}) as a perturbation, and use the
unperturbed normal modes, but correct the frequencies to first order:
$\omega_{\mu}' = \omega_{\mu} - \sum_i s_i\,u^i_{\mu\mu}/\omega_{\mu}$.
[Given the stated assumptions about the relative sizes of
the three terms in Eq.~(\ref{Htot}), the frequency shift can
indeed be seen to be small.]
With this approximation, the kernel $R_{\{s,s'\}}$
factorizes into a product
$\prod_{\mu}R_{\mu}$ of kernels for each mode (we suppress the Ising
variables where no confusion is possible), with
\begin{equation}
R_{\mu}(q'_{\mu}, q_{\mu})
       = \langle q'_{\mu}|e^{iH_{\mu}(\{s'\})\,t}
           e^{-iH_{\mu}(\{s\})\,t} |q_{\mu}\rangle,
\label{Rmu}
\end{equation}
and
$H_{\mu} = p_{\mu}^2/2m + m{\omega_{\mu}'}^2 q^2_{\mu}/2$.
Equation~(\ref{Rmu}) has a simple physical interpretation. Starting from
an initial state, the system evolves forward in time for a duration $t$
as a harmonic oscillator of frequency $\omega_1$, say. It then evolves 
backward in time for duration $t$ as a harmonic oscillator of slightly
different frequency
$\omega_2$. For our problem, this difference propagator, $R_{\mu}$,
can be 
further simplified because the frequencies $\omega_1$ and $\omega_2$ are
almost identical \cite{unn}. If we think
about the corresponding classical problem
in phase space, the forward and backward evolutions take place
on ellipses of nearly equal eccentricity. To good approximation, we may
regard the ellipses as coincident. With suitably scaled $p$ and $q$ axes,
this common ellipse is a circle, on which the particle
sweeps out angles $\omega_1 t$ and $-\omega_2 t\,$ in the forward and
backward motion. The net evolution is that of
a single harmonic oscillator of frequency $\delta_{\mu}=\omega_2 -
\omega_1$, and mass $m\omega_{\mu}/\delta_{\mu}$,
for a time $t$ \cite{expl}.  In other words, 
\begin{equation}
R_{\mu}(q'_{\mu},q_{\mu}) \approx \langle q'_{\mu} |
     e^{i (p^2_{\mu} + m^2\omega^2_{\mu}q^2_{\mu})\delta_{\mu}t
         /2m\omega_{\mu}} |q_{\mu}\rangle.
\label{Rmu2}
\end{equation}

It is now easy to carry out the coordinate integrals in
Eq.~(\ref{Qint2}) for the special case
where $\rho$ is a thermal equilibrium density matrix
$\propto e^{-\beta H_{\rm nm}}$ with $\beta=1/kT$. Since $\rho$
and $R$ both factorize by normal mode, i.e.,
$\rho = \prod_{\mu} \rho_{\mu}$, and
$R=\prod_{\mu}R_{\mu}$, the summand in Eq.~(\ref{Qint2}) also
factorizes into $\prod_{\mu} \Lambda_{\mu}(t)$, where
$\Lambda_{\mu}=\int\!\int R_{\mu}\rho_{\mu}dq_{\mu}dq'_{\mu}$.
By using standard coordinate representations
of the harmonic oscillator density matrix and propagator,
$\Lambda_{\mu}$ is easily evaluated, and the result can be written as
(restoring $\hbar$)
\begin{eqnarray}
P(t) &=  & 2^{-2N} \sum_{\{s,s'\}} \prod_{\mu}\Lambda_{\mu}(t),
     \label{ptfn}\\
\Lambda_{\mu}(t) & = & {\sinh\beta\hbar\omega_{\mu}/2
                       \over
                  \sinh(\beta\hbar\omega_{\mu} - i\delta_{\mu}t)/2}.
     \label{Lamb}
\end{eqnarray}
Note that $\delta_{\mu}= \sum_i (s_i -s'_i)u^i_{\mu\mu}/\omega_{\mu}$
depends on the Ising configuration, and $u^i_{\mu\mu}$ is given
below Eq.~(\ref{Ker}).

Equations (\ref{ptfn}) and (\ref{Lamb}) formally answer the question
we set out to investigate, but the sum on the $s$'s is nontrivial.
Some general properties of the result are worth noting, however.
Thus, $P(t)$ is real,
and since $|\Lambda_{\mu}(t)| \leq 1$, $P(t) \leq 1$. [In fact
$P(t) =1$ only if $\sum_i u^i_{\mu\mu}t/\pi\omega_{\mu}$ is an
integer for all $\mu$ simultaneously.] The expressions simplify greatly
at $T=0$. Then, $\Lambda_{\mu}(t) = e^{i\delta_{\mu}t/2}$, and
\begin{eqnarray}
P(t) & =  & \prod_i \cos^2(\zeta^2_i t/2\omega_0),
    \quad\quad (T=0),
         \label{Teq0} \\
\zeta^2_i & =  & \sum_{\mu} \zeta_{i\mu}^2 =
             \sum_{\mu}{\bf c}_{i\mu}\cdot{\bf c}_{i\mu}/
              m \hbar \omega_{\mu}.
\label{zi}
\end{eqnarray}
Note that $\zeta_i$ and $\zeta_{i\mu}$ have dimensions of frequency.
We can also obtain the decoherence time $\tau_d$ for all $T$ by
examining the initial drop of $P(t)$ from unity. Writing 
$1- P(t) \approx (t/\tau_d)^2$, we obtain
\begin{equation}
{1\over {\tau^2_d}} = {1\over 4\omega_0^2}
\sum_i\left[
  \biggl( \sum_{\mu}
    \zeta^2_{i\mu}\coth x_{\mu} \biggr)^2
 +
\sum_{\mu} \zeta^4_{i\mu} {\rm cosech}^2 x_{\mu} \right],
\label{taud}
\end{equation}
where $x_{\mu} = \beta\hbar\omega_{\mu}/2$.
Note that $\tau_d$ falls as $T$ rises, as it should. At $T=0$,
$\tau_d^{-2} = \sum_i \zeta^4_i/4\omega^2_0$.

We still need the couplings ${\bf c}_{i\mu}$. These depend
on the nature of states $|e\rangle$ and $|g\rangle$,
so we will find them only for a particularly favorable situation
obtained by using Ba$^+$ ions, and states with $\Delta M = \pm 1$
in the 6s~$^2\!S_{1/2}$ and
5d~$^2\!D_{5/2}$ (or $^2\!D_{3/2}$) multiplets for $|g\rangle$
and $|e\rangle$ respectively. The $^2\!D_{5/2} \to {^2\!S}_{1/2}$ decay
is an E2 process, with $\tau_{\rm sp} \simeq 35$ s
\cite{Chu}, $\omega_0/2\pi = 1.7\times 10^{14}$~Hz.
The interaction hamiltonian is given by
\begin{equation}
H' = q\mathop{{\sum}'}_{\!i,j}\sum_{\alpha = x,y}
   z^{-4}_{ij} u^{ij}_{\alpha} {\hat Q}^i_{\alpha z},
\label{Hquad}
\end{equation}
where $q$ is the ionic charge, $z_i$, ${\hat Q}^i_{\alpha \beta}$,
and $u^i_{\alpha}$ are the
equilibrium position, quadrupole moment tensor, and displacement
from equilibrium
of the {\it i\/}th ion; $z_{ij}=z_i-z_j$, 
$u^{ij}_{\alpha} = u^i_{\alpha} - u^j_{\alpha}$.
Note that only transverse vibrations appear in Eq.~(\ref{Hquad})
because states $|e\rangle$ and $|g\rangle$ are connected by
$|\Delta M| = 1$; the same restriction ensures that the
equilibrium quadrupole fields do not drive any $|e\rangle
\leftrightarrow |g\rangle$ transitions. Writing the mode index
$\mu=(r,\alpha)$, with $r=1,2,\ldots,N$, we can write
$u^i_{\alpha} = \sum_r F^i_r q_{r\alpha}$, where the
$F^i_r$ are normal mode eigenvectors \cite{inpol}.
We scale these to obey $\sum_r F^i_r F^j_r = \delta_{ij}$,
$\sum_i F^i_r F^i_s = \delta_{rs}$. With
$F^{ij}_r\equiv F^i_r - F^j_r$, and
$Q^i_{\alpha\beta} \equiv \langle e| {\hat Q}^i_{\alpha\beta} | g\rangle$,
we obtain
$|{\bf c}_{i,r\alpha}| = 
     -q\sum'_j F^{ij}_r z^{-4}_{ij} |Q^i_{\alpha z}|$.

It remains to substitute the above
expression for ${\bf c}_{i,r\alpha}$
into Eqs.~(\ref{zi}) and (\ref{taud}) to obtain
$\tau_d$. Here we will only outline a simplified
calculation at $T=0$. The steps are as follows. First,
in Eq.~(\ref{zi}) we replace $\omega_{\mu}$ by
$\omega_{t,N}$, the frequency of the zigzag transverse mode, for
all $\mu$. Since $\omega_{t,N}$ is the
smallest normal mode frequency, this replacement yields a lower
bound for $\tau_d$. It also obviates finding the $F^i_r$,
as the sum over $r$ can be done by
orthonormality of the $F^i_r$. Second, for the states in
question $|Q^i_{\alpha z}|^2 = 18\hbar/k_0^5 \tau_{\rm sp}$, with
$k_0 = \omega_0/c$, and $\alpha = x,y$.
Third, we note that if $\omega_z$ is
the longitudinal CM vibrational frequency,
$d_0=(q^2/m\omega^2_z)^{1/3}$ is a natural
trap length scale. We define the dimensionless sums
$S_n(i) = \sum'_j z_{ij}^{-n}d^n_0$. [The expression for
$\tau^{-2}_d$ contains a factor
$\sum_i \left ( S_4^2(i) + S_8(i) \right )^2.$]
We estimate $S_n(i)$ using a
continuum approximation for the ion array \cite{Dub}. In this
approximation, 
the local interionic spacing is $s(z) = s(0)(1-z^2/L^2)^{-1}$, where
$s(0)=4L/3N$, and $2L$ is the total length of the array,
with $L^3 \approx 3N\ln(0.8N)$. (All lengths are in units of $d_0$.)
This yields $S_n(i)\approx 2\zeta(n)/s^n(z_i)$. The sum over $i$ can
now be estimated by an integral.
Combining these results, we obtain a bound for $\tau_d$ entirely
in terms of trap and ion parameters,
\begin{equation}
{1\over \tau_d} <  0.36 {1\over \tau_{\rm sp}}
      {N^{35/6} \over [\ln(0.8 N)]^{8/3}}
      {\omega^2_z \over \omega_0 \omega_{t,N} (k_0d_0)^5}.
\label{taulb}
\end{equation}
The last step is to estimate $\omega_{t,N}$.
By considering the transverse force on the central
ion, we get $\omega_{t,N}^2 \approx \omega_t^2 - c(N)\omega^2_z$,
where $\omega_t$ is the CM transverse vibrational frequency, and
$c(N) = 9\zeta(3)N^2/ 16\ln(0.8N)$. In fact, requiring
$\omega_{t,N} > 0$ gives the critical $\omega_t$ ($\omega_{t,{\rm cr}}$)
needed to avoid the zigzag instability \cite{Dub}.
The numerical value of $\tau_d$ implied by Eq.~(\ref{taulb}) is a
very sensitive function of $\omega_z$ ($\sim \omega_z^{-16/3}$).
By choosing $\omega_z/2\pi$ in the 10--100 kHz range, and
$\omega_{t,N} \simeq \omega_{t,{\rm cr}}/2$, the ratio
$\tau_d/\tau_{\rm sp}$ can be seen to lie in the $10^4$--$10^8$ range
for $N=1000$. [The assumptions behind Eqs.~(\ref{phase}),
(\ref{Rmu}), and (\ref{Rmu2}), can all be seen to hold.]

We thus see that contrary to what might have been expected,
vibrations of the ions are not a significant source of
decoherence in the CZ QC for $N\leq 1000$. Larger $N$ values pose serious
technical challenges in trap design and in keeping $s(0)$ large enough
to be optically resolvable. The conclusions 
of the present paper, however, can only be encouraging for the prospect
of quantum computation.

I am grateful to H.~J. Kimble and D.~Wineland for useful comments 
and correspondence. This work is supported by NSF grant no.
DMR-9306947.


\begin{references}
\bibitem{lat} A longer paper, with many refinements, will be published
elsewhere.
\bibitem{unn} This simplification is not essential, but the
formula for $\Lambda_{\mu}(t)$ [see Eq.~(\ref{Lamb})]
is very complex without it.
\bibitem{expl} If we regard Eq.~(\ref{Rmu})
(dropping the subscript ${\mu}$)
as the matrix element of a unitary operator ${\hat R}(t)$, and define
$q(t) = {\hat R}^{\dagger}(t) q {\hat R}(t)$, and likewise $p(t)$, the
linearity of harmonic oscillators guarantees that $q(t)$ and $p(t)$ are
the same formal functions of $q(0)$ and $p(0)$ as the solution
to the classical problem. The argument in the text shows that in
replacing Eq.~(\ref{Rmu}) by Eq.~(\ref{Rmu2}), the error
is of relative order $\delta_{\mu}/\omega_{\mu}$ for all $t$.
Hence, the errors in the mean and the width of the position and
momentum of a quantum mechanical
wavepacket are similarly small for all $t$. One can find
complicated initial and final states for which the
approximation~(\ref{Rmu2}) is poor, but these are not contained in the
thermal density matrix $\rho(Q,Q')$ which we employ.
\bibitem{inpol} For the linear Paul trap as in Ref.~\cite{Rai}, $F^i_r$
is independent of $\alpha$, including $\alpha = z$.
\end{references}
\end{document}